\documentclass[twocolumn,notitlepage,prb,color,superscriptaddress,psfig,showpacs,amsmath,amssymb,nobibnotes,longbibliography]{revtex4-2}
\usepackage[colorlinks=true,citecolor=blue]{hyperref}%

\usepackage{epsf}
\usepackage{graphicx}
\usepackage{amssymb}
\usepackage{color}
\usepackage{float}
\usepackage{epstopdf}
\usepackage{natbib}
\usepackage{amsmath}
\usepackage{mathrsfs}
\usepackage{xspace}
\usepackage{wrapfig}
\usepackage{soul}
\setstcolor{red}
\usepackage{verbatim}

\newcommand{\CGT}{Cr$_{\text{2}}$Ge$_{\text{2}}$Te$_{\text{6}}$\xspace}
\newcommand{\MnPS}{Mn$_2$P$_2$S$_6$\xspace}
\newcommand{\MnNiPS}{MnNiP$_2$S$_6$\xspace}
\newcommand{\NiPS}{Ni$_2$P$_2$S$_6$\xspace}
\newcommand{\PS}{(Mn$_{1-x}$Ni$_x$)$_2$P$_2$S$_6$\xspace}
\newcommand{\HIIab}{$\textbf{H} \parallel ab$\xspace}
\newcommand{\Hpab}{$\textbf{H}\,\perp\,ab$\xspace}

\newcommand{\md}[1]{{\color{black}{#1}}}
\definecolor{darkgreen}{rgb}{0, 0.4, 0}

\usepackage{soul}
\usepackage[normalem]{ulem}

\definecolor{mygray}{cmyk}{0, 0, 0, 0.3}

\begin{document}
\title{Evolution of the spin dynamics in the van der Waals system $\boldsymbol{M}_{\text{2}}$P$_{\text{2}}$S$_{\text{6}}$ ($\boldsymbol{M}_{\text{2}}$ = Mn$_{\text{2}}$, MnNi, Ni$_{\text{2}}$) series probed by electron spin resonance spectroscopy}

\author{Y.~Senyk}
\thanks{These authors contributed equally to this work.}
\affiliation{Leibniz IFW Dresden, D-01069 Dresden, Germany}
\affiliation{Institute for Solid State and Materials Physics, TU Dresden, D-01062 Dresden, Germany}
\author{J.~J.~Abraham}
\thanks{These authors contributed equally to this work.}
\affiliation{Leibniz IFW Dresden, D-01069 Dresden, Germany}
\affiliation{Institute for Solid State and Materials Physics, TU Dresden, D-01062 Dresden, Germany}
\author{Y.~Shemerliuk}
\affiliation{Leibniz IFW Dresden, D-01069 Dresden, Germany}
\author{S.~Selter}
\affiliation{Leibniz IFW Dresden, D-01069 Dresden, Germany}
\affiliation{Institute for Solid State and Materials Physics, TU Dresden, D-01062 Dresden, Germany}
\author{S.~Aswartham}
\affiliation{Leibniz IFW Dresden, D-01069 Dresden, Germany}
\author{B.~B\"uchner}
\affiliation{Leibniz IFW Dresden, D-01069 Dresden, Germany}
\affiliation{Institute for Solid State and Materials Physics and W{\"u}rzburg-Dresden Cluster of Excellence ct.qmat, TU Dresden, D-01062 Dresden, Germany}
\author{V.~Kataev}
\affiliation{Leibniz IFW Dresden, D-01069 Dresden, Germany}
\author{A.~Alfonsov}
\affiliation{Leibniz IFW Dresden, D-01069 Dresden, Germany}
\affiliation{Institute for Solid State and Materials Physics and W{\"u}rzburg-Dresden Cluster of Excellence ct.qmat, TU Dresden, D-01062 Dresden, Germany}

\begin{abstract}
		
	In this work we report a detailed ESR spectroscopic study of the single-crystalline samples of the van der Waals compounds $\boldsymbol{M}_{\text{2}}$P$_{\text{2}}$S$_{\text{6}}$ ($\boldsymbol{M}_{\text{2}}$ = Mn$_{\text{2}}$, MnNi, Ni$_{\text{2}}$), performed at an excitation frequency of 9.56 GHz, in a broad range of temperatures above the magnetic order, and at different orientations of the magnetic field with respect to the sample. Analyzing temperature and angular dependences of the resonance field and of the linewidth of the \MnPS compound we have observed a significant change of the spin dynamics from the dominance of the 3D-like fluctuations close to the magnetic order to a relative increase of the 2D-like spin fluctuations at higher temperatures. Such a behavior, which is opposite to the development of the low-D signatures in the previously studied \CGT compound, can be explained by the difference in the type of magnetic order in \MnPS and \CGT. On the other hand, \MnNiPS compound exhibits angular dependences of the linewidth typical for the system with 3D-like spin correlations in the whole measurement temperature range, however the 2D-like correlations can be seen in the temperature dependences of the resonance field and the linewidth. \NiPS, in turn, does not show any 2D signatures. This suggests that varying the Ni content in \PS one can control the exchange interaction, possibly also in the third dimension. 
		
\end{abstract}

\date{\today}
\maketitle

\section{Introduction}

Within the recent years layered magnetic van der Waals (vdW) materials attracted considerable attention in the field of fundamental science as well as in the applied research due to their natural tendency to exhibit a low dimensional behavior. By virtue of the vdW interaction which does not build up chemical bonds and thus only weakly couples atomic layers together, even in a bulk vdW material such low dimensionality is often strongly pronounced in physical properties \cite{Pokrovsky90}. This enables these materials to demonstrate new fundamental physical phenomena and, therefore, opens a possibility for their application in next-generation spintronic devices \cite{Huang2017,Gong2017,Otrokov2019,Gong2019,Yang21}.

Particularly interesting subfamily of these compounds is represented by the transition metal (TM) tiophosphates $M_{\text{2}}$P$_{\text{2}}$S$_{\text{6}}$ where $M$ stands for a TM ion \cite{Brec86,Grasso02}. Here the crystallographic $ab$-plane hosts the spins of $M$ ions, which are arranged on a two-dimensional (2D) honeycomb spin lattice \cite{Wang2018}. Due to the high flexibility of the choice of the $M$ ion, it is possible to control the magnetic properties, which, in turn,  allows realization of different Hamiltonians in the $M_{\text{2}}$P$_{\text{2}}$S$_{\text{6}}$ family. For instance, if a system with $M$\,=\,Mn can be described by the Heisenberg antiferromagnetic exchange with an easy-axis magnetic anisotropy term \cite{Joy92,Wildes98, Shemerliuk2021, Lu2022, Abraham2022}, which stabilize the antiferromagnetic order at $T_N \approx 77$\,K, then in the case of $M$\,=\,Fe the antiferromagnetism is found to be of the Ising type \cite{Joy92,Lancon16,Selter21}. A compound with $M$\,=\,Ni exhibits a more complex behavior, which possibly can be explained by the anisotropy of the Heisenberg exchange of the {\it XXZ} type \cite{Joy92,Wildes15,Lancon18,Selter21,Shemerliuk2021,Lu2022}. In this case the magnetic order sets in at $T_N \approx 158$\,K, which is the highest in this series. Interestingly, in our previous work we have shown that \NiPS exhibits a 3D-like spin dynamics rather then 2D-like in both magnetically ordered and paramagnetic states, which could be the result of a significant interlayer coupling \cite{Mehlawat2022}. Since low-dimensionality in layered vdW materials is the key property, the study of its signatures and understanding of its dependence on the sample composition are very important. 

In this work we present the results of a detailed X-band ($\nu = 9.56$\,GHz) electron spin resonance (ESR) spectroscopic study of single crystalline samples of \MnPS, \MnNiPS and \NiPS carried out in a broad range of temperatures above the magnetic order. The ESR method, enabling a direct access to the electronic spin system, was recently successfully applied for studies of several magnetic van der Waals compounds providing important insights into the peculiarities of their spin dynamics \cite{Okuda1986,joy1993,Kobets2009,Otrokov2019,Zeisner2019,Khan2019,Saiz2019,Vidal2019,Zeisner2020,Saiz2021,Singamaneni2020,Ni2021,Alahmed2021,Sakurai2021,Alfonsov2021,Alfonsov2021b}. Here we show that signatures of the low dimensionality, seen in the characteristic angular dependence of the ESR linewidth, are present only in the \MnPS compound. Interestingly, this material shows a development of the 2D behavior upon increasing temperature starting from the temperature $T \sim 150$\,K and up to the room temperature, which is opposite to the development of the low-D signatures in \CGT \cite{Zeisner2019}, which is also a layered van der Waals magnet, however, with the ferromagnetic order stabilizing at a temperature $T_C \approx 61 - 66$\,K. The \MnNiPS compound exhibits angular dependences of the linewidth typical for the system with 3D-like spin correlations in the whole measurement temperature range, however the 2D-like correlations can be seen in the temperature dependences of the resonance field and the linewidth. \NiPS, in turn, does not show any 2D signatures. Bearing in mind the sensitivity of the characteristic angular dependence of the linewidth to the short-wavelength antiferromagnetic fluctuations, our finding suggests that an increase of the Ni content in \PS increases the exchange interaction, which is possibly extended to the third dimension, i.e., along the direction normal to the crystallographic $ab$-plane.

\section{Experimental Details}
\label{Sec:ExpDet}

Single crystals of \MnPS, \MnNiPS and \NiPS \md{ studied in this work} were grown using chemical vapor transport method. \md{Details of their growth, crystallographic, compositional and static magnetic characterization are} described in Refs. \cite{Selter21, Shemerliuk2021}. \md{Note, that the experimental value $x_{exp}$ in \PS for the nominal \MnNiPS compound is found to be $x_{exp} = 0.45$, considering an uncertainty of approximately 5\% \cite{Shemerliuk2021}.}  The ESR measurements were carried out at a microwave frequency of $\nu$ = 9.56\,GHz using Bruker X-band ESR-spectrometer. 
%
%\md{ on the samples from the same batch as those reported in \cite{Shemerliuk2021}}. 
%
The magnetic field $H$ was swept from 0 to 9\,kOe. Samples were placed in a $^4$He-gas flow cryostat (Oxford Instruments) allowing temperature measurements between 4 and 300\,K. The experimental setup was equipped with a goniometer for angular dependent measurements. All the recorded spectra are field derivatives of the microwave power absorption. 

The measured spectra consist of a single line which have been fitted with the field derivative of the Dysonian function \cite{Feher1955,Dyson1955}:
\begin{eqnarray}
&&\frac{dD}{dH}  =  \frac{dD_{+}}{dH} + \frac{dD_{-}}{dH}; \; \;  x_{\mp}  = 2 \frac{H \mp H_{res}}{\Delta H}; \nonumber\\
&&\frac{dD_{\mp}}{dH}  \propto  \frac{2}{\Delta H} \left(\frac{1-x_{\mp}^2}{(1+x_{\mp}^2)^2} \sin(\phi)  - \frac{2x_{\mp}}{(1+x_{\mp}^2)^2} \cos(\phi) \right). \; \; \; \; \;
\label{eq:lorentz}
\end{eqnarray}

Here, $D$ is the Dysonian line shape function where $\mp H_{res}$ is the resonance field, negative or positive, $\Delta H$ is the full width at half maximum. The first term in brackets in Eq.~(\ref{eq:lorentz}) is the dispersion part of the spectra and the second term is the absorption part of the spectra. The contribution  with the negative resonance field $\frac{dD_{-}}{dH}$ is necessary to correctly fit broad lines. Parameter $\phi$ represents the instrumental mixing effect of absorption and dispersion parts. In the cases of \MnPS and \MnNiPS a very good fit of the Eq.~(\ref{eq:lorentz}) to the measured data showed that the mixing was negligible ($\phi=0$). Therefore, the line shape was Lorentzian, i.e., given by the pure absorption term in Eq.~(\ref{eq:lorentz}). From the fit, $\Delta H$ and $H_{res}$ values were obtained. 

For a precise alignment of the sample in the magnetic field parallel to the $ab$-plane and perpendicular to the $ab$-plane, the angular dependence of the resonance field $H_{res}(\theta)$ and the linewidth $\Delta H(\theta)$ was used.

\section{Theoretical framework}
\label{sec:TF}

In the canonical theories describing the implications of the electron spin dynamics in low-dimensional magnets on the characteristics of the ESR response, that are reviewed, e.g., in Ref.~\cite{Benner1990}, the spin system is parameterized by the Hamiltonian comprising the Zeeman interaction $\mathcal{H}_{\rm Z} $, the isotropic Heisenberg exchange $\mathcal{H}_{\rm iso}$, and the weaker anisotropic magnetic couplings such as the dipole-dipole (d-d) $\mathcal{H}_{\rm d-d}$ interaction and the symmetric anisotropic exchange (pseudo-dipolar)  interaction $\mathcal{H}_{\rm aniso}$:

\begin{eqnarray}
	\mathcal{H}_{\rm Z} & = &g\mu_{\rm B}\sum_{\rm i }\mathbf{H}\cdot \mathbf{S}_{\rm i}\ ; \label{eq:Zeeman}\\
	\mathcal{H}_{\rm iso} & = & J\sum_{\rm i \neq j} \mathbf{S}_{\rm i}\cdot \mathbf{S}_{\rm j}\ ; \label{eq:iso_exch}\\
	\mathcal{H}_{\rm d-d} & = & \sum_{\rm i < j} g^2\mu_{\rm B}^2\left[  \frac{\mathbf{S}_{\rm i}\cdot \mathbf{S}_{\rm j}}{r^3_{\rm ij}}   - 
	\frac{3(\mathbf{S}_{\rm i}\cdot \mathbf{r}_{\rm ij})(\mathbf{S}_{\rm j}\cdot \mathbf{r}_{\rm ij})}{r_{\rm ij}^5}\right]\,; \label{eq:dip_dip}\\
	\mathcal{H}_{\rm aniso} & = & \sum_{\rm i \neq j} \mathbf{S}_{\rm i}\cdot A \cdot \mathbf{S}_{\rm j}\ . \label{eq:aniso_exch}
\end{eqnarray}   
Here, $J$ is the isotropic exchange constant and $A$ is the tensor of the symmetric anisotropic exchange. Under the ESR resonance conditions the circular polarized microwaves of the frequency $\omega_0$ couple to the transverse magnetization $M_+$ of the electron spin ensemble precessing at the Larmor frequency $\omega_{\rm L} = \omega_0$  which causes absorption of electromagnetic energy. If only isotropic interactions (\ref{eq:Zeeman})  and (\ref{eq:iso_exch}) were present, the absorption line would be a delta function since both $\mathcal{H}_{\rm Z} $ and $\mathcal{H}_{\rm iso} $   commute with $M_+$, i.e., $M_+$ is a conserved quantity under these conditions. In contrast, anisotropic interactions (\ref{eq:dip_dip}) and (\ref{eq:aniso_exch}) do not commute with $M_+$ causing broadening of the ESR signal. Usually its width can be treated as a sum of two contributions \cite{Huber1972,Seehra1972}: 
\begin{eqnarray}
	\Delta H & = & \Delta H(T) + \Delta H(T \rightarrow \infty)\,. \label{eq:linewidth_total}
\end{eqnarray}
$\Delta H(T \rightarrow \infty)$ in Eq.~(\ref{eq:linewidth_total})  stands for the angle and temperature independent part corresponding to the linewidth in the infinite temperature limit, whereas $\Delta H(T)$ is the temperature and angular dependent contribution. It is a product of two terms, the  $T$-dependent and isotropic part which incorporates the spin dynamics and the angular part which reflects the symmetry of the spin lattice and the symmetry of the interactions. 

Specifically, assuming a dipolar-like symmetry of the anisotropic part of the interaction, $\Delta H(T)$ can be expressed in terms of the spectral densities of the spin correlation functions $\tilde{\Gamma}(\omega)$ at low frequencies $\omega < J/\hbar$ as \cite{Benner1990}:
\begin{eqnarray}
	\Delta H(T) & = & \omega_{\rm d}^2[F_0^2\tilde{\Gamma}(0) + 10 F_1^2\tilde{\Gamma}(\omega_0) + F_2^2\tilde{\Gamma}(2\omega_0)]\,. \label{eq:linewidth_Tdep}
\end{eqnarray}  
Here, $\omega_{\rm d}$ is the dipolar coupling constant, the first term in brackets is the contribution from the secular part of the d-d interaction
that connects the $S_{\rm i}$ and $S_{\rm j}$  states for which the total magnetization $M_{\rm z} = \sum_{\rm i}S_{\rm z}$ does not change  ($\Delta M_{z} = 0$), while the second and the third terms are due to the nonsecular part of the d~-~d interaction with $\Delta M_{z} = \pm 1$ and $\Delta M_{z} = \pm 2$, respectively, \cite{Slichter96} all weighted by the respective geometrical coefficients $F_{0,1,2}$~\cite{Richards1974}.

\begin{figure}[t]
	\centering
	\includegraphics[width=\columnwidth]{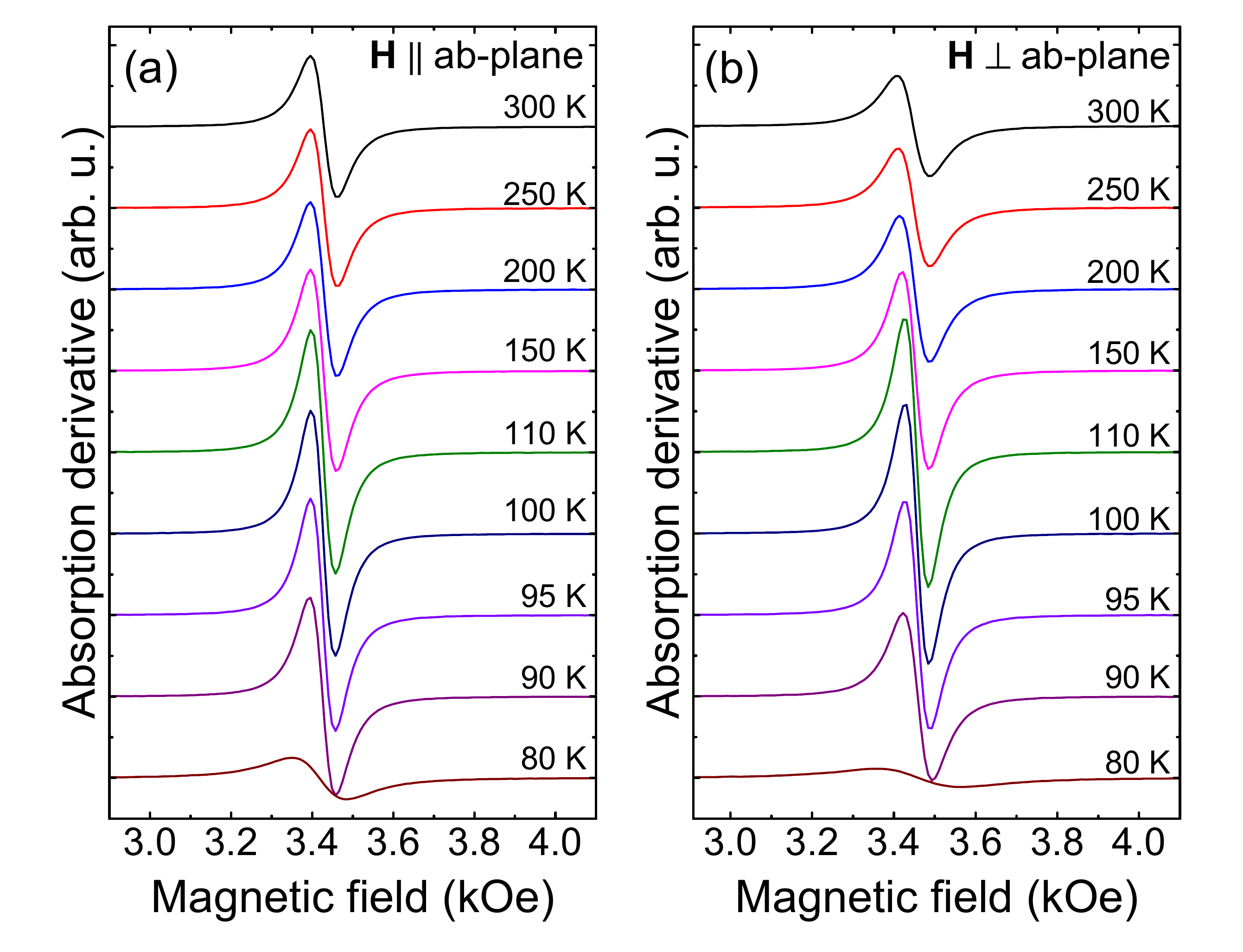}
	%\captionsetup{justification=Justified}
	\caption {Temperature dependence of the \MnPS ESR spectra at a fixed excitation frequency $\nu$ = 9.56\,GHz for \HIIab-plane (a) and \Hpab-plane (b).}
	\label{fig:spectra1}
\end{figure}

The specific type of the angular part of Eq.~(\ref{eq:linewidth_Tdep}) depends on the details of the spin fluctuation spectrum. In the layered spin systems with strong interlayer couplings the spin self-correlation function decays rapidly on the timescale of $J^{-1}\hbar$. At such short times spin fluctuation modes with all wave vectors $q$ contribute to the fluctuation spectrum and summing the geometrical coefficients in Eq.~(\ref{eq:linewidth_Tdep}) over all wave vectors yields the angular dependence of the type $\Delta H(T) \propto (\cos^2\theta+1)$, where $\theta$ is the angle between the field vector $\mathbf{H}$ and the normal to the spin plane \cite{Huber1972,Richards1974}. If the spin planes are decoupled, correlations can retain for much longer time manifesting in a long diffusive tale of the spin self-correlation function at $t > J^{-1}\hbar$ \md{for both FM and AFM type of intra-plane exchange}. Since the low $q$ modes have the longest lifetimes they dominate in this regime. For $q \rightarrow 0$ the geometrical coefficients in Eq.~(\ref{eq:linewidth_Tdep}) simplify to \cite{Richards1974,Benner1978}:
\begin{eqnarray}
	F_0^2 & = & (3\cos^2\theta - 1)^2\ ,\\
	F_1^2 & = & \sin^2\theta \cos^2\theta\ , \\
	F_2^2 & = & \sin^4\theta\ . \label{eq:F_coeff}
\end{eqnarray}
Since the secular part of the d-d interaction makes the dominant contribution to the linewidth (first term in Eq.~(\ref{eq:linewidth_Tdep})), its angular dependence acquires roughly the form  $\Delta H(T) \propto (3\cos^2\theta-1)^2$ \cite{Richards1974}.

If the spins interact predominantly antiferromagnetically, lowering the temperature would result in a 3D long-range AFM order of the layered spin system at a finite $T_{\rm N}$ due to the residual interlayer coupling, or the substantial magnetic anisotropy. An enhancement of the short-wavelength AFM spin fluctuations at the ordering $q$-vector by approaching $T_{\rm N}$ on expense of the long-wavelength ($q \rightarrow 0$)  fluctuations would result in the change of the type of the angular dependence from $(3\cos^2\theta-1)^2$ towards $(\cos^2\theta+1)$ \cite{Richards1974}. In contrast, in the case of the ferromagnetic interaction, the $(3\cos^2\theta-1)^2$ type of dependence may become even more pronounced by approaching the magnetic phase transition due to the enhancement of the long-wavelength FM spin fluctuations.   

\begin{figure}[t]
	\centering
	\includegraphics[width=\columnwidth]{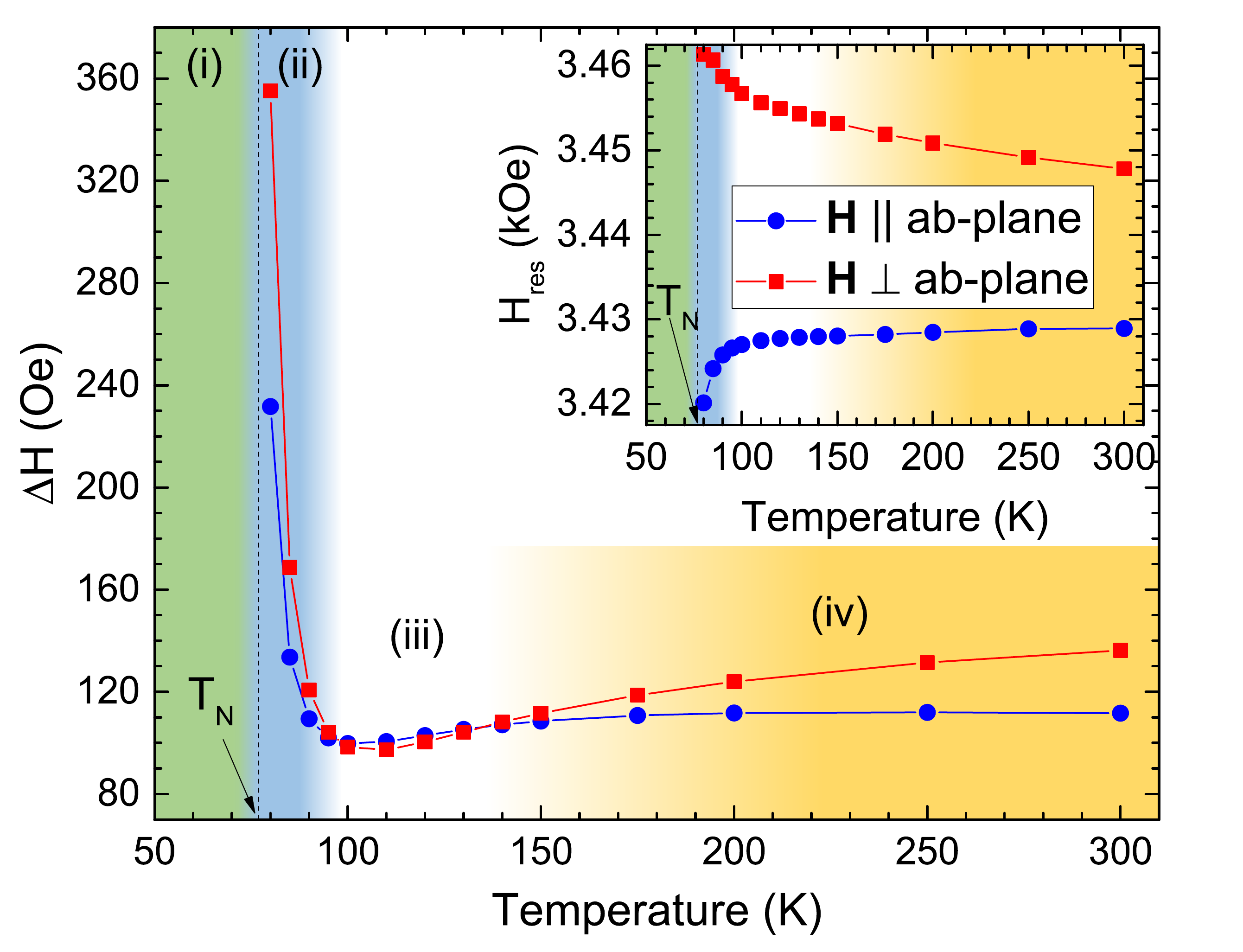}
	%\captionsetup{justification=Justified}
	\caption{\MnPS : Temperature dependence of the linewidth at $\nu$ = 9.56\,GHz for \HIIab-plane (closed circles) and \Hpab-plane (closed squares). The respective temperature dependence of the resonance field is shown in the inset. Differently colored temperature ranges indicate four regimes of the spin dynamics discussed in Sec.~\ref{sec:MnPS}.}
	\label{fig:temp1}
\end{figure}

\section{Experimental Results}
\subsection{\MnPS}
\label{sec:MnPS}

X-band spectra of \MnPS at various temperatures for magnetic field applied parallel and perpendicular to the crystallographic $ab$-plane are shown in Fig.~\ref{fig:spectra1}(a) and Fig.~\ref{fig:spectra1}(b), respectively. As can be seen in the figures, approaching the ordering temperature $T_N\sim77$\,K \cite{Shemerliuk2021} the intensity of the spectra drastically reduces and the linewidth increases due to the growing correlations and the opening of the antiferromagnetic (AFM) excitation gap at low temperatures \cite{Okuda1986, Kobets2009, Abraham2022}, as expected for the magnetically ordered materials. The disappearance of the ESR signal in this case can be used for the estimation of $T_N$.

	\begin{figure}[t]
	\centering
	\includegraphics[width=\columnwidth]{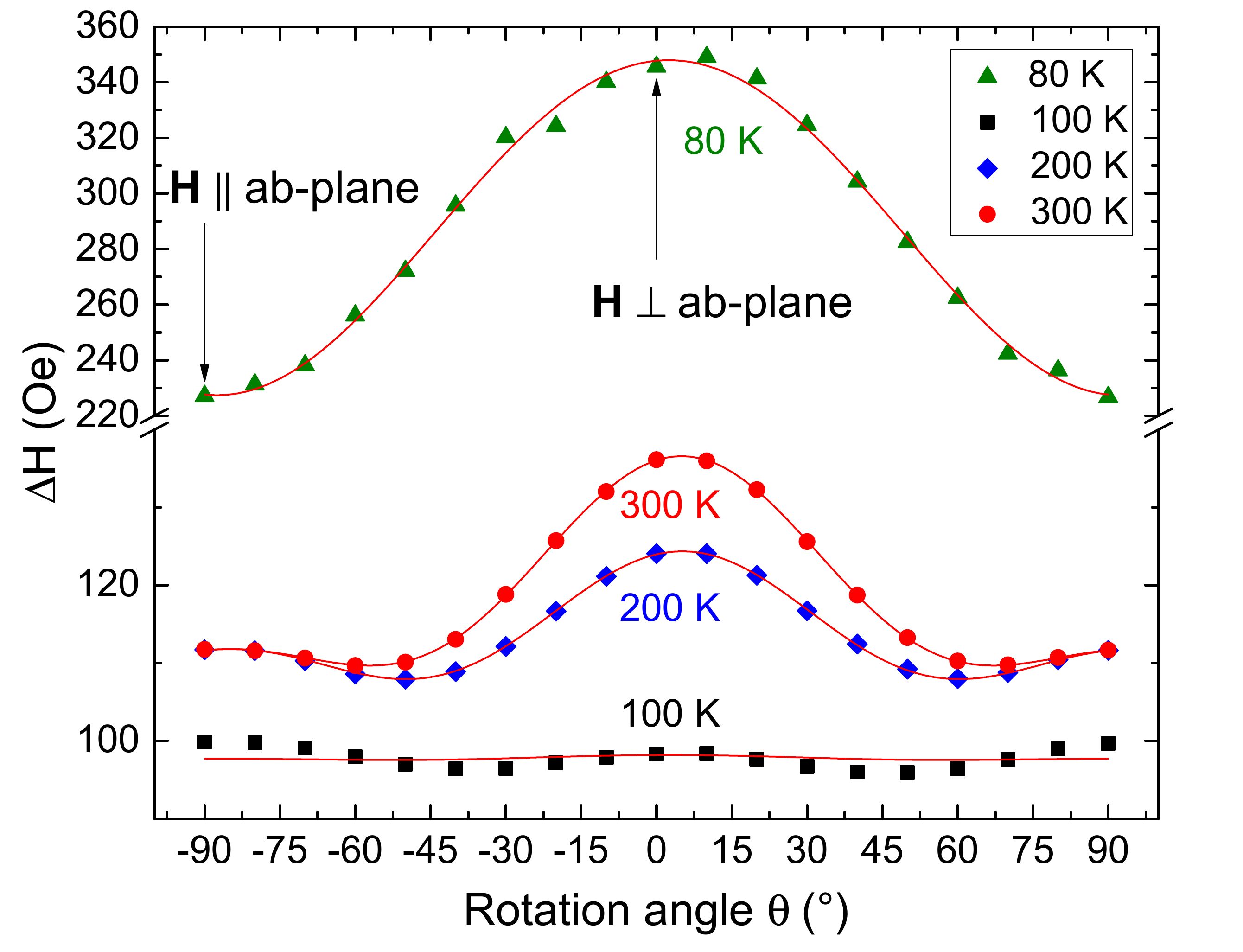}
	%\captionsetup{justification=Justified}
	\caption{\MnPS : Angular dependence of the linewidth at $\nu$ = 9.56\,GHz at four different temperatures : 80\,K, 100\,K, 200\,K and 300\,K. Solid red lines are fits to the data as described in the text.}
	\label{fig:angular1}
	\end{figure}

In Fig.~\ref{fig:temp1}, the temperature dependence of the linewidth and the resonance field, obtained according to Sec.~\ref{Sec:ExpDet} (Eq.~(\ref{eq:lorentz})), are shown for both orientations \HIIab-plane and \Hpab-plane. With decreasing temperature the $H_{res}$ value for \HIIab orientation (inset in Fig.~\ref{fig:temp1}) remains practically constant down to a temperature of about 100\,K, then there is a shift of the line due to the development of the AFM spin correlations, static on the ESR time scale. For the \Hpab configuration the line shift is visible in the whole measurement temperature range. As for the linewidth, for the \Hpab orientation it decreases with lowering the temperature down to approximately 100\,K and below this temperature it starts strongly increasing towards the ordering temperature. For the \HIIab orientation, above 200\,K the linewidth remains constant. With further lowering the temperature it slightly decreases down to approximately 100\,K and demonstrates  an increase below 100\,K similar to the \Hpab orientation.

In order to get insights on the linewidth anisotropy at different temperatures (Fig.\,\ref{fig:temp1}) we have measured angular dependence of the ESR spectra at various temperatures. The obtained dependences of the linewidths are presented in Fig.~\ref{fig:angular1}. At a temperature of 80\,K which is close to the AFM transition, $\Delta H(\theta)$ is proportional to the (cos$^{2}(\theta)+1)$ dependence, where $\theta$ is the angle between the applied magnetic field and the direction normal to the $ab$-plane\md{, i.e., the spin plane}. As explained in Sec.~\ref{sec:TF}, such an angular dependence is typical for spin systems arranged in three-dimensional lattices with a noticeable anisotropic coupling \cite{Richards1974,Benner1990}, or in two-dimensional lattices with a significant interlayer coupling, which renders them quasi-three-dimensional. At a temperature of 100\,K the $\Delta H(\theta)$ dependence is very weak. At higher temperatures of 200 and 300\,K $\Delta H(\theta)$ has a characteristic minimum close to the angle of $\theta$\,=\,$55^o$, which is a signature of the low-dimensionality of the spin system (Sec.~\ref{sec:TF}). These two dependences can be fitted with additional contribution of the form proportional to (3\,cos$^{2}(\theta)-1)^{2}$, \md{which is accounted for in the phenomenological} total \md{angular dependence of the} linewidth $\Delta H(\theta) = C_2\,($\,cos$^{2}(\theta) + 1 ) + C_4\,(3\,$cos$^{2}(\theta)-1)^{2} + C_0$ \md{(solid lines in Fig.~\ref{fig:angular1})}. \md{Note, that this \md{simplified} equation does not provide \md{precise weights} of different contributions, but rather serves as an indicator of the prominence of the 2D correlations.} \md{As a result, we observe, that at} 300\,K the contribution to the angular dependence of the linewidth due to low-dimensionality is stronger than that at 200\,K. \md{Interestingly, such an evolution of the angular dependence of the ESR linewidth with changing temperature was observed in \MnPS compound by \citet{Okuda1986}, which supports our finding.}

Thus, \md{the $\Delta H(T, \theta)$ dependences presented in Figs.\, \ref{fig:temp1} and \ref{fig:angular1} enable} to determine four regimes of the spin dynamics highlighted in Fig.~\ref{fig:temp1}: (i) $T<T_N$ -- long-range magnetic order; (ii) $T_N<T<\sim100$\,K -- strong dominance of the quasi-static on the ESR time scale correlations of a 3D-like character; (iii) $\sim100$\,K $<T<$ $\sim150$\,K \md{almost} isotropic regime, representing a crossover from 3D- to 2D-like dynamics; (iv) $T>\sim150$\,K growing \md{prominence} of the correlations of a 2D-like character.

	\begin{figure}[t]
	\centering
	\includegraphics[width=\columnwidth]{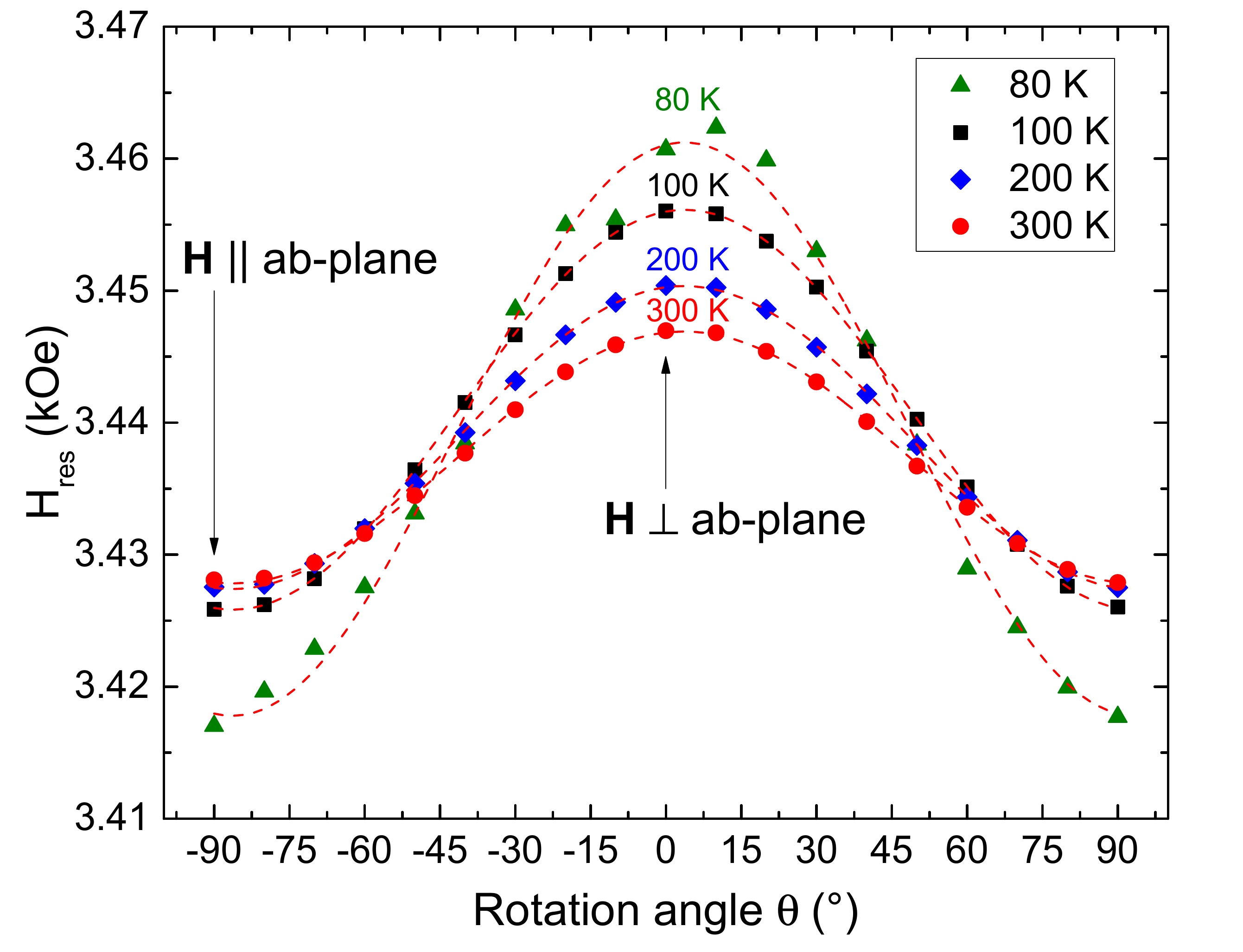}
	%\captionsetup{justification=Justified}
	\caption{\MnPS : Angular dependence of the resonance field $H_{res}$ at $\nu$ = 9.56\,GHz at four different temperatures : 80\,K, 100\,K, 200\,K and 300\,K. Dashed red lines are fits to the data as described in the text.}
	\label{fig:rf1}
	\end{figure}

In addition to the angular dependence of the linewidth $\Delta H(\theta)$, the angular dependence of $H_{res}$ at various temperatures was measured. At all temperatures it is given by $H_{res}(\theta) \propto \cos^{2}(\theta)$ (Fig.~\ref{fig:rf1}). Interestingly, even at the highest measurement temperature, the resonance fields still show a pronounced angular dependence, which suggests a g-factor anisotropy \cite{Abraham2022} and a presence of static on the ESR time scale local magnetic fields due to 2D correlations seen in the angular dependence of the linewidth at these temperatures. The latter is also suggested by the gradual increase of the resonance field with decreasing temperature measured for the \Hpab configuration (inset in Fig.~\ref{fig:temp1}).

\subsection{\MnNiPS}

\begin{figure}[t]
	\centering
	\includegraphics[width=\columnwidth]{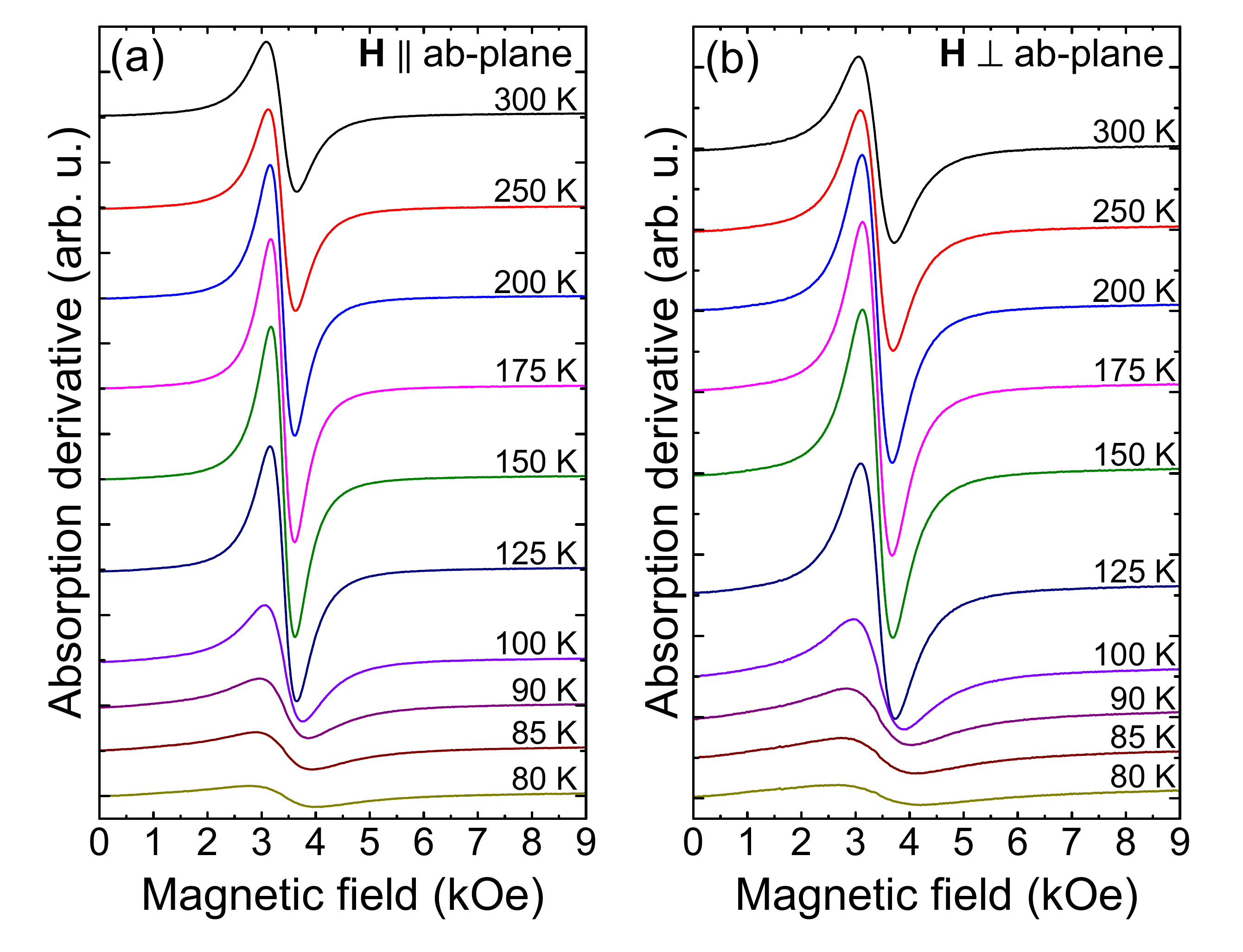}
	%\captionsetup{justification=Justified}
	\caption{Temperature dependence of the \MnNiPS ESR signal at a fixed excitation frequency $\nu$ = 9.56\,GHz for \HIIab-plane (a) and \Hpab-plane (b).}
	\label{fig:spectra2}
	\end{figure}

Characteristic X-band ESR spectra of \MnNiPS for a wide range of temperatures are presented for both \HIIab and \Hpab orientations in Fig.~\ref{fig:spectra2}(a) and Fig.~\ref{fig:spectra2}(b), respectively. Similarly to the \MnPS sample, the temperature of the vanishing of the ESR signal can be used for the estimation of the ordering temperature for this particular sample. As a result we get a value of \md{$T_N$, which is equal to or less than $\sim 60$\,K}. \md{It agrees well with the value of $T_N \sim 57$\,K obtained from the measurement of the temperature dependence of the susceptibility $\chi$ at $H = 1000$\,Oe for the ${\bf H} \perp ab$ configuration, performed on the same sample} \footnote{\md{Note, that various studies have reported different values of $T_N$ for \MnNiPS compound, for instance it amounts to 12\,K in} \cite{Basnet2021}, 38\,K in \cite{Shemerliuk2021}, 41\,K in \cite{Yan2011} and 42\,K in \cite{Lu2022}. \md{Such a variation in $T_N$ could be due to a stochastic distribution of Mn and Ni ions on the 4g Wyckoff sites of the crystal structure, and due to small variations of the real Mn/Ni content.}}.

	\begin{figure}[t]
	\centering
	\includegraphics[width=\columnwidth]{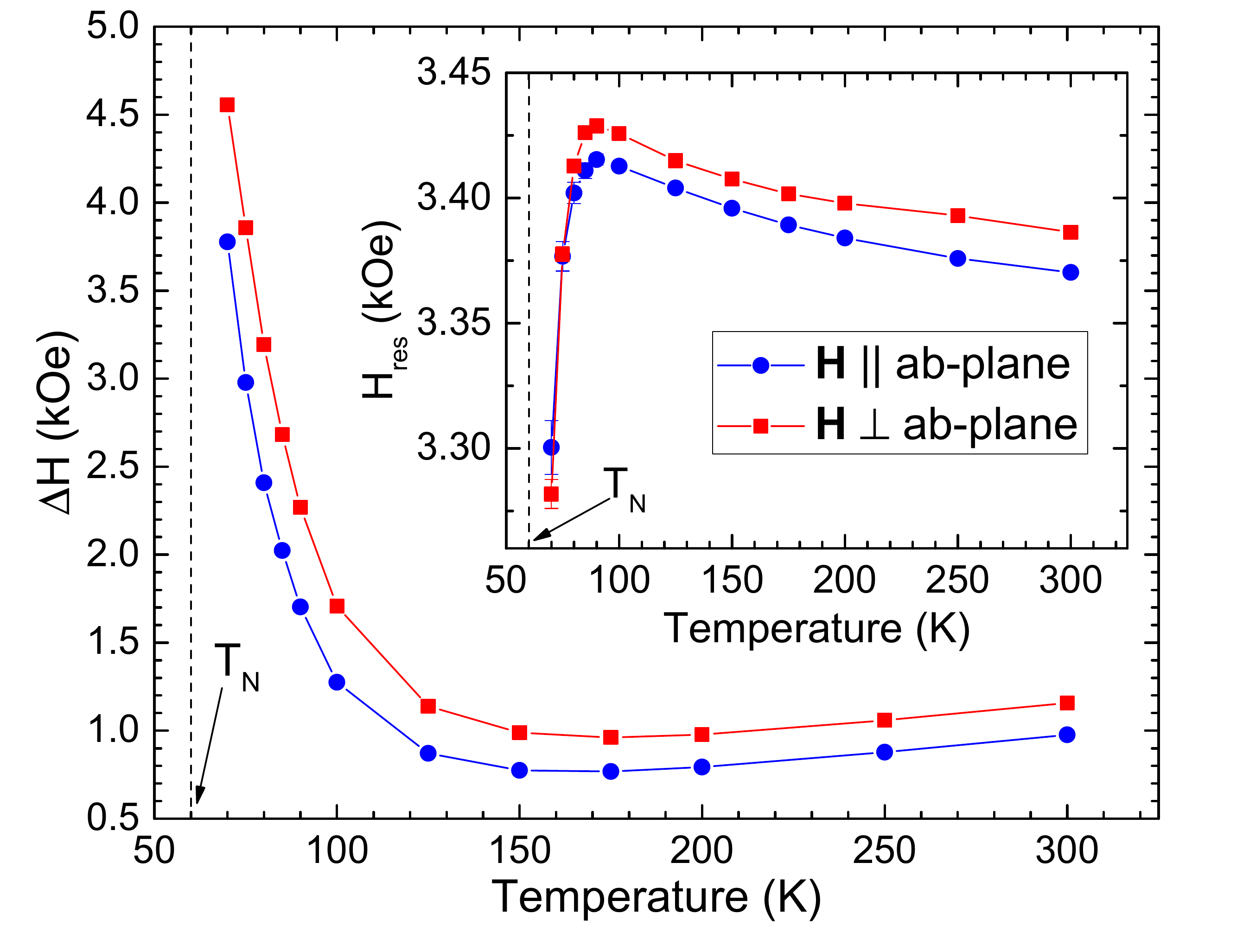}
	%\captionsetup{justification=Justified}
	\caption{\MnNiPS : Temperature dependence of the linewidth at $\nu$ = 9.56\,GHz for \HIIab-plane (blue circles) and \Hpab-plane (red squares). The respective temperature dependence of the resonance field is shown in the inset.}
	\label{fig:temp2}
	\end{figure}	

	\begin{figure}[t]
	\centering
	\includegraphics[width=\columnwidth]{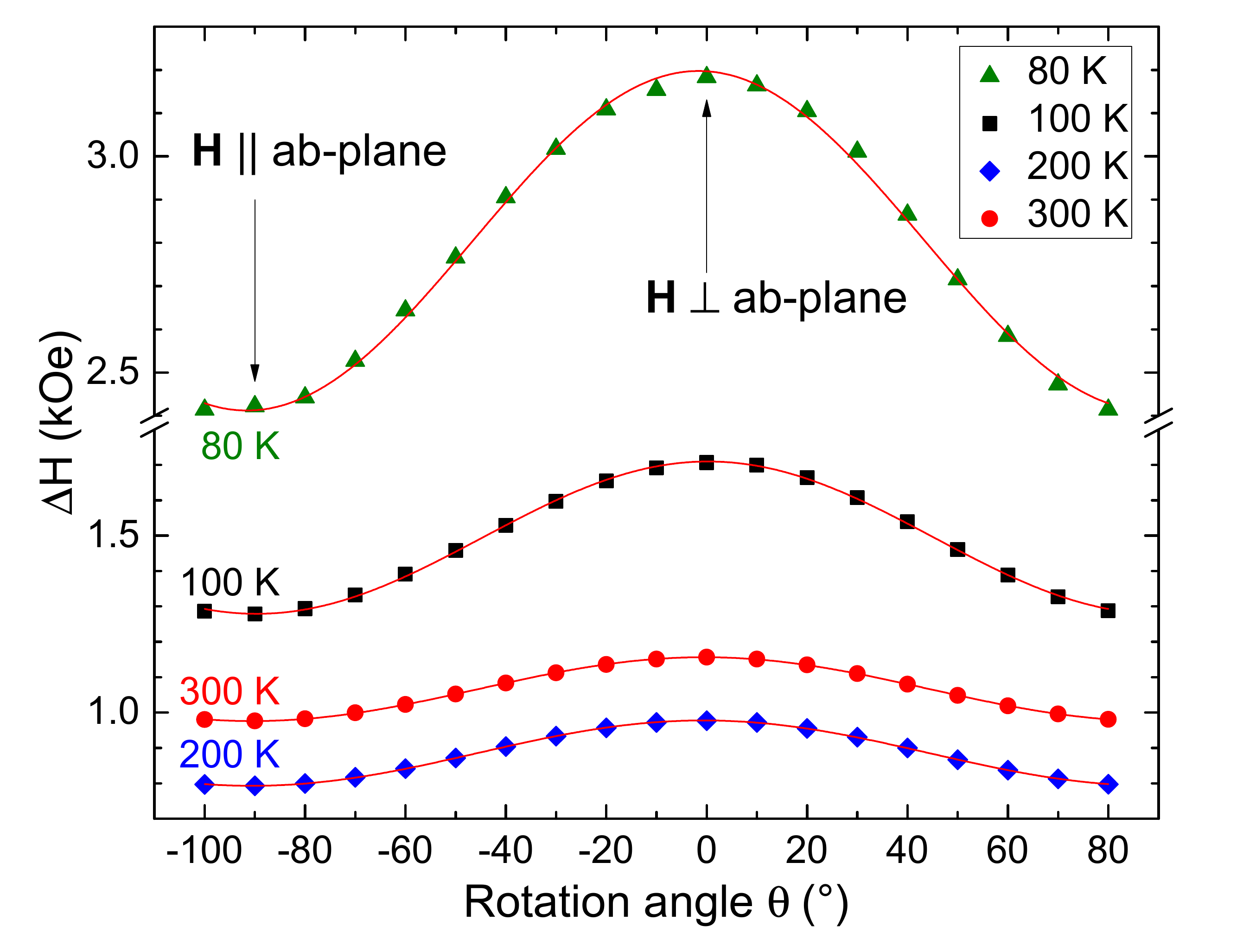}
	%\captionsetup{justification=Justified}
	\caption{\MnNiPS : Angular dependence of the linewidth at $\nu$ = 9.56\,GHz at four temperatures : 80\,K, 100\,K, 200\,K and 300\,K. Solid red lines are fits to the data as described in the text.}
	\label{fig:angular2}
	\end{figure}

The linewidth evolution with temperature for two magnetic field orientations observed for the \MnNiPS compound is somewhat different from the one observed for \MnPS (Fig.~\ref{fig:temp2}). A shallow minimum for both orientations \HIIab and \Hpab occurs at $T\sim175$\,K, which is higher than that for \MnPS. Upon further reduction of temperature the linewidth increases likely due to the growing 2D correlations. Note that the linewidth at each temperature is several times larger than in the case of the pure \MnPS compound. The inset of Fig.~\ref{fig:temp2} shows the temperature evolution of $H_{res}$ for \MnNiPS. The rapid decrease of $H_{res}$ before reaching the ordering temperature is related to the development of AFM spin correlations, static on the ESR time scale. At temperatures higher than ~90\,K, $H_{res}$ for both orientations of the magnetic field progressively changes as a function of temperature. This could be due to the static on the ESR timescale fields, resulting from the short range 2D correlations present up to the highest measurement temperature of 300\,K.

The angular dependence of the linewidth $\Delta H(\theta)$ for \MnNiPS reveals a pure (cos$^{2}(\theta)+1)$ type at all measured temperatures (Fig.~\ref{fig:angular2}). This suggests a strong dominance of the three-dimensional spin correlations seen in the ESR linewidth in the whole measurement temperature range, which is in contrast to the \MnPS compound. Since \md{nevertheless,} signatures of the two-dimensionality are observed in the temperature dependence of the resonance field and the linewidth, the absence of the characteristic (3\,cos$^{2}(\theta)-1)^{2}$ dependence in the angular dependence of the linewidth is possibly due to an increased isotropic exchange interaction (Sec.~\ref{sec:TF}), which could be induced by the Ni ions present in \MnNiPS.

\begin{figure}[t]
	\centering
	\includegraphics[width=\columnwidth]{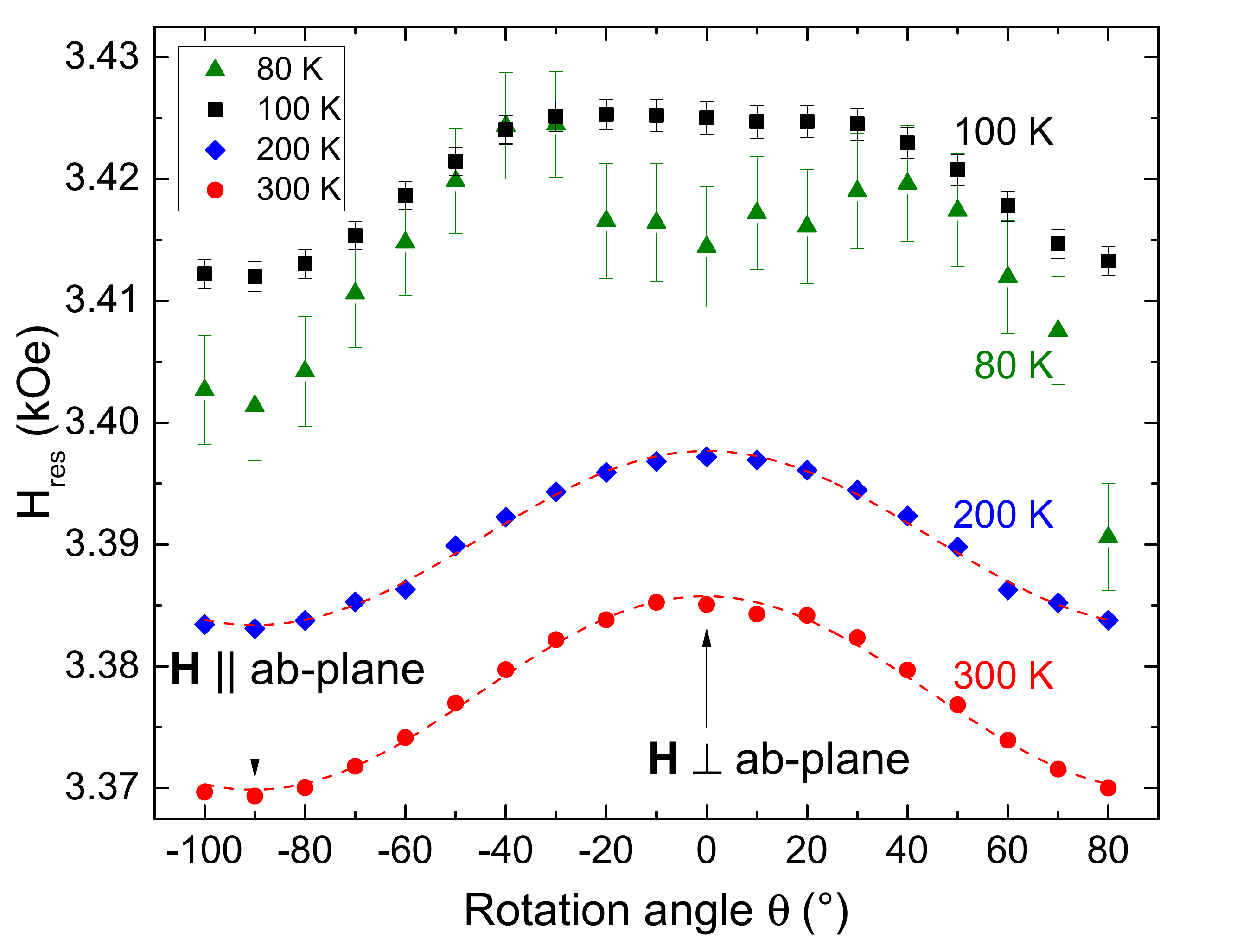}
	%\captionsetup{justification=Justified}
	\caption{\MnNiPS : Angular dependence of the resonance field $H_{res}$ at $\nu$ = 9.56\,GHz at four different temperatures : 80\,K, 100\,K, 200\,K and 300\,K. Dashed red lines are fits to the data as described in the text.}
	\label{fig:rf2}
	\end{figure}

The angular dependences of $H_{res}$ for \MnNiPS at four temperatures are presented in Fig.~\ref{fig:rf2}. At higher temperatures of 300 and 200\,K they \md{they can be well fitted by the} $H_{res}(\theta) \propto \cos^{2}(\theta)$ dependence, whereas at lower temperatures of 100 and 80\,K the $H_{res}(\theta)$ dependences progressively deviate from $\sim\cos^{2}(\theta)$, suggesting a more complicated type of magnetic order possessing a symmetry which is different from an easy-axis biaxial AFM in \MnPS.

\subsection{\NiPS}

\begin{figure}[t]
   	\centering
   	\includegraphics[width=\columnwidth]{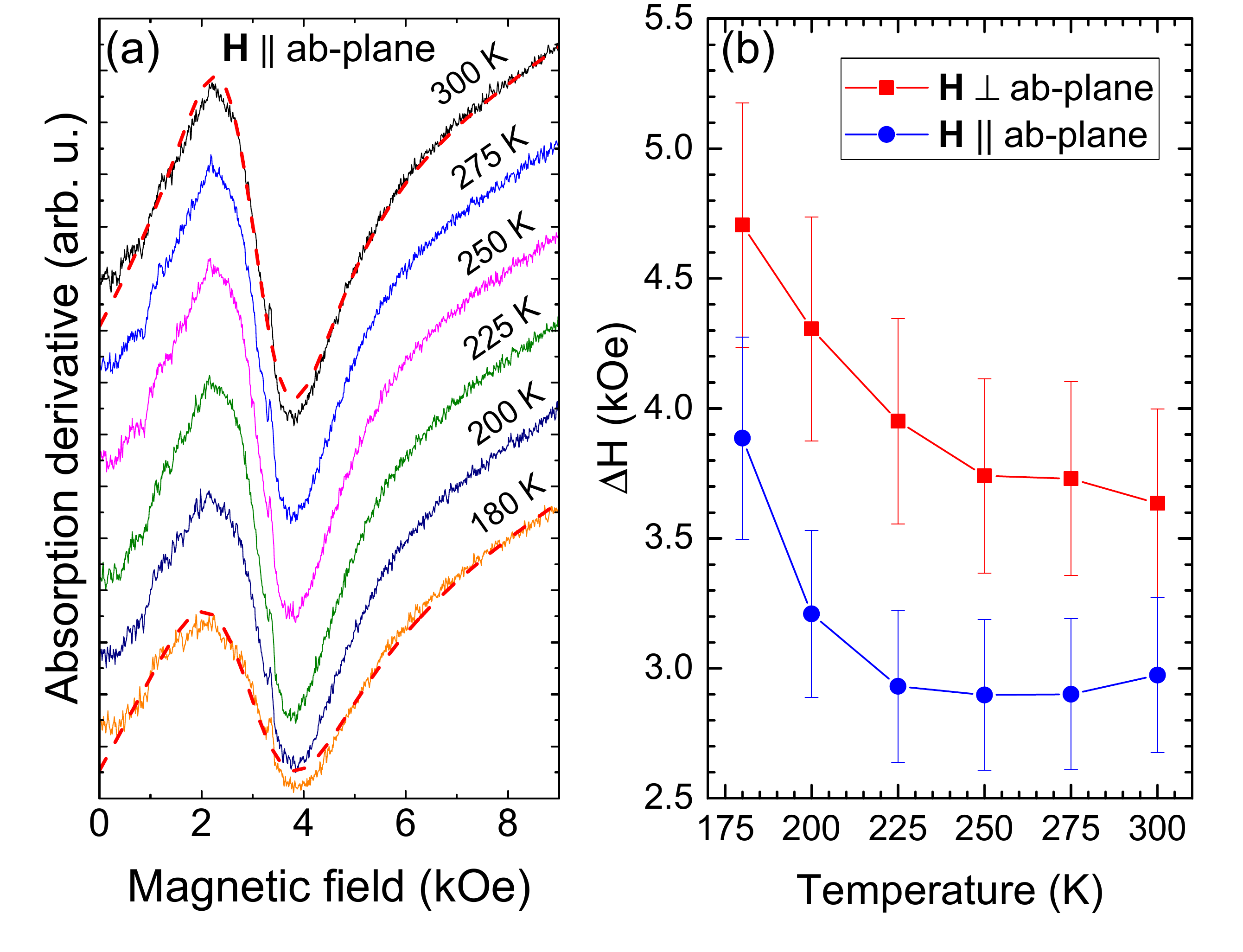}
   	%\captionsetup{justification=Justified}
   	\caption {\NiPS: a) Temperature dependence of the \NiPS ESR signal at a fixed excitation frequency $\nu$ = 9.56\,GHz for \textbf{H}\,$||$\,\textit{ab}-plane orientation. Dashed red lines are the fits using the function described in Eq.(\ref{eq:lorentz}). b) Temperature dependence of the linewidth at $\nu$ = 9.56\,GHz for \HIIab-plane (blue circles) and \Hpab-plane (red squares). }
   	\label{fig:nips}
\end{figure}

X-band ESR spectra of \NiPS were measured for both \HIIab (Fig.~\ref{fig:nips}(a)) and \Hpab (not shown) orientations at temperatures above $T_{N}$. The spectra were fitted with the function given in Eq.~(\ref{eq:lorentz}). In contrast to the \MnPS and \MnNiPS cases, here the obtained fits show some deviations from the measured spectra. The deviations are resulting likely from the extremely small intensities and the large linewidths of the ESR lines in the case of the \NiPS sample. The linewidth at any temperature is larger than that in \MnPS and \MnNiPS compounds. The small intensity increases the signal to noise ratio, and the linewidth, that is larger than the resonance field results in the violation of the resonance condition, which might cause the distortion of the line shape. Nevertheless, the dependence of the linewidth as a function of temperature resulting from the fit can be used for the comparison with the other samples, however, one has to bear in mind large error bars. As can be seen in Fig.~\ref{fig:nips}(b), close to $T_N$ the ESR linewidth of \NiPS increases with decreasing temperature, which is visible for both magnetic field orientations, \HIIab and \Hpab. Taking into account the error bars, the difference between the linewidth measured for different directions of the magnetic field is quite small, which, together with results of the study of the high frequency/field ESR on this compound \cite{Mehlawat2022}, allows to conclude on the absence of the signatures of 2D-like spin fluctuations in \NiPS compound. 

\section{Discussion}

Comparing \MnPS, \MnNiPS and \NiPS samples, first of all, it is found that the ESR linewidth in the paramagnetic state above $T_N$ progressively increases with increasing the Ni content. In contrast to Mn$^{2+}$ with the half filled 3d electronic shell, and a small admixture of the excited state $^4 P _{5/2}$ into the ground state $^6 S _{5/2}$, the ground state of the Ni$^{2+}$ ion in the octahedral environment \cite{Wang2018} is a spin triplet with the higher lying orbital multiplets, admixed through the spin-orbit coupling \cite{AbragamBleaney}. Such a second-order spin-orbit coupling effect enhances a sensitivity of the Ni spin ($S = 1$) to the small differences in the local environment. As has been shown in Ref.~\cite{Mehlawat2022}, due to the sizable interaction of Ni spins, causing the exchange narrowing effect, variation of the local environment should yield a change of the effective g-factor value. The disorder, resulting from stochastic mixing of two magnetically inequivalent ions in \MnNiPS compound, therefore, could explain the increase of the linewidth. However, if such disorder would play a major role in the determination of the linewidth, then the largest linewidth should be observed in the \MnNiPS compound, which is not the case. Another possible source of the broadening of the line is the anisotropy of the exchange interaction (Sec.~\ref{sec:TF}). In the case of symmetric anisotropic exchange \md{(Eq.~\ref{eq:aniso_exch})}, the linewidth is proportional to $A^2 / (\vert J \vert)$ \cite{Kubo1954,Kataev2001}, where $A \propto (\Delta g / g)^2 J $ \cite{Moriya1960}, yielding:
\begin{eqnarray}
\label{eq:dHgJ}
\Delta H \propto \left(\frac{\Delta g}{g}\right)^4 J 
\end{eqnarray}
Here \md{$\Delta g = |g-2|$} is the deviation of the \md{average} g-factor from the free electron spin only value \md{caused by the spin-orbit coupling}.
According to Ref.~\cite{Mehlawat2022}, in the \NiPS compound $\Delta g$ amounts to $\sim 0.166$. In \MnPS one can try to estimate $\Delta g$ from the resonance fields for \HIIab and \Hpab at 300\,K, assuming that the shift of the resonance field is given only by the spin-orbit coupling effect, which likely would give an overestimated value. In this case for \MnPS $\Delta g$ is about $\sim 0.016$\md{, which is comparable to the values found in other Mn doped tiophosphates \cite{Lifshitz1982,Sibley1994}}. Similar estimation of the g-factor \md{shift} for \MnNiPS compound yields a value of $~0.019$. \md{The observed} general trend of increasing $\Delta g$ with increasing Ni content suggests that the anisotropy of the exchange interaction might have a substantial contribution to the linewidth in the paramagnetic state \md{[Eq.~(\ref{eq:dHgJ})]}.

Interestingly, this is not the only effect that Ni substitution has on the spin dynamics of the title compounds. In the \NiPS compound three-dimensionality \md{of the spin lattice} is suggested by the small anisotropy of the linewidth and by the results reported in Ref.~\cite{Mehlawat2022}. The \MnNiPS compound does not show signatures of the 2D-correlations in the angular dependences of the linewidth in the whole measurement temperature range, \md{although} such correlations can be seen in the temperature dependence of the resonance field and the linewidth. The pure Mn compound \MnPS shows signatures of the 2D-like spin dynamics in the angular dependence of the linewidth. However, they become visible only at high temperatures, i.e., far away from the ordering temperature. This, first, suggests that the magnetic order itself is of a 3D nature, provided by the magnetic anisotropy and the non-zero inter-plane interaction, which was also observed in, e.g., layered vdW magnetic topological insulators \cite{Alfonsov2021b}. In order to suppress this interaction, it is apparently necessary to increase the temperature. This behavior is opposite to the one, which is found in another representative of magnetic vdW materials, \CGT \cite{Zeisner2019}. There an enhancement of the 2D-like spin fluctuations was observed upon decreasing the measurement temperature, and was found strongest in the paramagnetic state close to the magnetic ordering temperature. Such a contrasting behavior in \CGT and \MnPS is likely related to the difference in the type of magnetic order, being easy-axis ferromagnetic in the former \cite{Zeisner2019}, and easy-axis antiferromagnetic (with some anisotropy in the $ab$-plane) in the latter case \cite{Kobets2009, Abraham2022}, respectively. As is explained in Sec.~\ref{sec:TF}, in the case of a 3D long-range AFM order an enhancement of the short-wavelength AFM spin fluctuations at the ordering $q$-vector by approaching $T_{\rm N}$ on expense of the long-wavelength ($q \rightarrow 0$) fluctuations hinders the observation of the 2D spin dynamics in the angular dependence of the linewidth \cite{Richards1974}. In contrast, in the case of the ferromagnetic interaction, the 2D correlations may become even more pronounced by approaching the magnetic phase transition due to the enhancement of the long-wavelength FM spin fluctuations.

\md{
Previously, the critical behavior of \MnPS both in the magnetically ordered state at $T < T_{\rm N}$ and also above $T_{\rm N}$ was studied by nuclear magnetic resonance (NMR) spectroscopy \cite{Berthier1978,Torre1989,Dioguardi20,Bougamha2022} and neutron scattering \cite{Wildes98,Wildes2006,Wildes2007}. Regarding the paramagnetic state at $T > T_{\rm N}$, which is in the focus of the present work, the static critical behavior probed by the temperature dependent  measurements of the $^{31}$P NMR Knight shift \cite{Dioguardi20,Bougamha2022} and by the analysis of the magnetic correlation length by the neutron scattering \cite{Wildes2006} suggests a 2D character of the spin correlations in \MnPS. This is also supported by the observation of the broad maximum of the static susceptibility at $T \approx 118 \text{\,K} > T_N$ \cite{Shemerliuk2021}, expected for the 2D systems \cite{Benner1990,DeJongh2001}. 
%Note, that such a \imp{broad} maximum of the susceptibility \imp{well} above the magnetic ordering temperature is \out{practically absent} \imp{not observed} in the \MnNiPS compound \cite{Shemerliuk2021}.
The dynamic critical behavior assessed by the analysis of the temperature dependence of the $^{31}$P spin-lattice relaxation in \MnPS implied the mean-field behavior. On the other hand, the dynamical critical neutron scattering indicates that the critical dynamics is completely confined to the $ab$~planes above $T\sim 105$\,K described at best with the 2D anisotropic Heisenberg model with a crossover to the 3D Heisenberg behavior by approaching $T_{\rm N}$ \cite{Wildes2006}. This is consistent with the above discussed angular dependence of the ESR linewidth in \MnPS\ showing a qualitatively similar crossover. 
}

\md{The absence of such} 2D signatures in the \md{angular dependence of the} ESR linewidth of the \MnNiPS compound, and the absence of any 2D correlations in \NiPS suggest that \md{an increase of the} Ni content in \PS\ \md{enhances} the isotropic exchange interaction, possibly also in the third dimension, i.e., along the $c$-axis. \md{Indeed, a recent study of the magnetic dynamics of \NiPS with inelastic neutron scattering reveals from the analysis of the spin wave dispersion a substantially stronger interlayer exchange interaction as compared to \MnPS \cite{Wildes2022}.}
%This is also supported by the very close values of the estimated $\Delta g$ in \MnPS and \MnNiPS, suggesting a higher isotropic exchange constant $J$ in the mixed compound to account for an overall about ten times larger linewidth value (see Eq.~(\ref{eq:dHgJ})). 
\md{At first glance one would expect that} the increased exchange interaction as a function of Ni content should also increase the ordering temperature $T_N$ in \PS. However, this is not the case as the smallest $T_N$ is found in the \MnNiPS compound \cite{Shemerliuk2021, Lu2022}. \md{Such a suppression of $T_{\rm N}$ despite  a stronger magnetic exchange} could be explained by the increased disorder due to the stochastic mixing of the magnetically inequivalent Mn and Ni ions, and by the competition of two different types of order in pure \MnPS (\md{out-of-plane} N\'eel type AFM) and \NiPS (\md{in-plane} zigzag type AFM) \cite{Lu2022}.

\section{Conclusion}

In summary, we have performed a detailed ESR spectroscopic study of the single-crystalline samples of the van der Waals compounds \MnPS, \MnNiPS and \NiPS, the members of the family of TM tiophosphates. The measurements were carried out at an excitation frequency of 9.56 GHz, in a broad range of temperatures above the magnetic order, and at different orientations of the magnetic field with respect to the sample. By analyzing the temperature dependence of the resonance field and of the linewidth of the \MnPS compound we have determined several regimes of the spin dynamics. Interestingly, the angular dependence of the ESR linewidth measured at high temperatures $T>\sim 150$\,K showed signatures which are characteristic for the 2D spin correlations. In contrast to the previously studied magnetic vdW compound \CGT \cite{Zeisner2019}, \MnPS shows a development of the 2D behavior upon increasing temperature, which we attribute to the difference in the type of magnetic order in these compounds. The \MnNiPS compound\md{, on the other hand,} shows signatures of 2D correlations only in the temperature dependences of the resonance field and the linewidth. \NiPS, in turn, exhibits temperature and angular dependences of the linewidth typical for the system with 3D-like spin correlations in the whole measurement temperature range. Our study, therefore, suggests that an increase of the Ni content in \PS\ \md{suppresses the low-dimensional spin dynamics at temperatures above the magnetic order due to the} increase \md{of} the exchange interaction, possibly also in the third dimension. Understanding of the details of such \md{compositional dependence of the magnetic exchange} calls for theoretical modeling.

\section*{Anknowledgments}

This work was supported by the Deutsche Forschungsgemeinschaft (DFG) through grants No. KA 1694/12-1, AL 1771/8-1, AS 523/4-1, and within the Collaborative Research Center SFB 1143 ``Correlated Magnetism – From Frustration to Topology'' (project-id 247310070), and the Dresden-Würzburg Cluster of Excellence (EXC 2147) ``ct.qmat - Complexity and Topology in Quantum Matter'' (project-id 390858490), as well as by the UKRATOP-project (funded by BMBF with Grant No. 01DK18002).
%\imp{(Alexey: Please check project numbers)}.

%\section*{Appendix A: X-band ESR spectroscopy on N\lowercase{i}$_2$P$_2$S$_6$}

\bibliography{ESR_Xband_MnPS}	

\end{document}